\documentclass[a4paper,11pt]{article}
\pdfoutput=1 

\usepackage{jheppub} 
\usepackage[T1]{fontenc} 
\usepackage[dvipsnames]{xcolor}
\usepackage{amsmath}
\usepackage{amssymb}
\usepackage{amsfonts}
\usepackage{comment}
\usepackage{enumerate}
\usepackage{changes}
\usepackage{color}
\usepackage{multirow}
\usepackage{graphicx}
\usepackage{tablefootnote}

\definecolor{darkgray}{gray}{0.25}
\definecolor{darkgreen}{rgb}{0,0.5,0}

\newcommand{\beq}{\begin{equation}}
\newcommand{\eeq}{\end{equation}}

\def\be{\begin{equation}}
\def\ee{\end{equation}}
\def\bea{\begin{eqnarray}}
\def\eea{\end{eqnarray}}

\title{Probing axions with neutron star inspirals and other stellar processes}

\author[a,b]{Anson Hook,}
\emailAdd{hook@stanford.edu}
\author[a,c]{Junwu Huang}
\emailAdd{jhuang@perimeterinstitute.ca}

\affiliation[a]{Stanford Institute for Theoretical Physics, Stanford University, Stanford, CA 94305, USA}

\affiliation[b]{Maryland Center for Fundamental Physics, University of Maryland, College Park, MD 20742}

\affiliation[c]{Perimeter Institute for Theoretical Physics, Waterloo, Ontario N2L 2Y5, Canada}

\abstract{
In certain models of a QCD axion, finite density corrections to the axion potential can result in the axion being sourced by large dense objects.
There are a variety of ways to test this phenomenon, but perhaps the most surprising effect is that 
the axion can mediate forces between neutron stars that can be as strong as gravity.  These forces can be attractive or repulsive and their presence can
be detected by Advanced LIGO observations of neutron star inspirals.  By a numerical coincidence, axion forces between neutron stars with gravitational strength naturally have an associated length scale of tens of kilometers or longer, similar to that of a neutron star. Future observations of neutron star mergers in Advanced LIGO can probe many orders of magnitude of axion parameter space. Because the axion is only sourced by large dense objects, the axion force evades fifth force constraints. We also outline several other ways to probe this phenomenon using electromagnetic signals associated with compact objects.
}

\begin{document} 
\maketitle

\flushbottom

\section{Introduction}

Advanced LIGO recently discovered gravitational waves (GW) from black hole mergers~\cite{Abbott:2016blz,TheLIGOScientific:2016qqj,Abbott:2016nmj}.  As various upgrades come online, this discovery will soon turn into a torrent of new data on merging black holes and hopefully neutron stars, which  
will likely reveal new and exciting phenomena.  In this paper, we discuss how neutron star mergers as well as other astrophysical processes can be used to constrain various theories of QCD axions.

The QCD axion~\cite{Peccei:1977hh,Peccei:1977ur,Weinberg:1977ma,Wilczek:1977pj} was originally introduced as a solution to the strong CP problem and is one of the most minimal solutions.
The strong CP problem is the question why the neutron electric dipole moment (eDM) is so small.  The neutron eDM is proportional to
\bea
\overline \theta = \theta + \text{arg} \, \text{det} Y_u Y_d ,
\eea
where $\theta$ is the QCD theta angle and $Y_{u,d}$ are the up and down type Yukawas, respectively.  From limits on the neutron eDM, it is known that $\overline \theta \lesssim 10^{-10}$~\cite{Baker:2006ts}.
This bound is particularly confusing given that various elements of the Cabibbo-Kobayashi-Maskawa (CKM) matrix have been measured to have a large complex phase.

The axion solution to the strong CP problem consists of adding a new particle $a$ with the coupling
\bea
\frac{a}{f_a} \frac{g_s^2}{32 \pi^2} G^{\mu \nu} \tilde G^{\mu \nu},
\eea
where $g_s$ is the strong coupling constant, $G^{\mu \nu}$ is the gluon field strength and $\tilde G^{\mu \nu} = \frac{1}{2} \epsilon^{\mu \nu \rho \sigma} G_{\rho \sigma}$.  The above coupling dynamically sets
the neutron eDM to zero.  At low energies, the axion obtains a potential, with a period of $2\pi f_a$, from the coupling to gluons~\cite{diCortona:2015ldu}
\bea
\label{Eq: potential}
V \approx - m_\pi^2 f_\pi^2 \sqrt{1 - \frac{4 m_u m_d}{\left(m_u + m_d\right)^2} \sin^2 \left ( \frac{a}{2 f_a} \right )}.
\eea
Other contributions to the axion potential can be simply added to obtain the total axion potential.

Various experiments have been proposed to search for the QCD axion and axion like particles (ALPs) through its couplings to nucleons~\cite{Budker:2013hfa,Graham:2013gfa,Arvanitaki:2014dfa}, electrons and photons~\cite{Kahn:2016aff,Armengaud:2014gea,Asztalos:2009yp,Sikivie:1983ip,Shokair:2014rna,Ehret:2009sq}. Many constraints on the axion parameter space have since been obtained through both direct measurement with ground-based experiments~\cite{Anastassopoulos:2017ftl,Hagmann:1998cb,Abel:2017rtm} as well as indirect measurement of energy loss and energy transport in various astrophysical objects, for instance, SN1987~\cite{Raffelt:2006cw} (see the axion chapter in~\cite{Olive:2016xmw} and references therein for more details).

In this paper, we will consider a type of QCD axion that couples to nuclear matter and has a mass that is smaller than expected from equation~\ref{Eq: potential}, which the CASPER experiment~\cite{Budker:2013hfa,Graham:2013gfa} is searching for and which is discussed in~\cite{Blum:2014vsa}. Therefore, we are considering a QCD axion whose potential is~\footnote{The exact form of the potential does not have to match the potential in equation~\ref{Eq: potential} as long as the period of the potential is $2\pi f_a$.}
\bea
\label{Eq: our potential}
V = - m_\pi^2 f_\pi^2 \epsilon \sqrt{1 - \frac{4 m_u m_d}{\left(m_u + m_d\right)^2} \sin^2 \left ( \frac{a}{2 f_a} \right )}.
\eea

These axions are lighter than expected by a factor of $\sqrt{\epsilon}$ and necessarily have $f_a \gtrsim 4 \times10^8$ GeV~\cite{Raffelt:2006cw} due to supernova cooling arguments.  This translates to a mass $m_a \lesssim \sqrt{\epsilon} \, 16$ meV or $1/m_a \gtrsim 10^{-8}\, \text{km}/\sqrt{\epsilon}$.

Surprisingly, these types of QCD axions can mediate large forces between neutron stars that are not excluded by current fifth force measurements.
Not only that, but the more weakly coupled the axion is, the stronger the force becomes!
As an illustration, consider a non-relativistic gas of neutrons and protons.  The finite density corrections to equation~\ref{Eq: our potential} are reviewed in appendix~\ref{App: finite}.  The finite density potential for the axion is
\bea
\label{Eq: finite density}
V = - m_\pi^2 f_\pi^2 \left\{\left(\epsilon - \frac{\sigma_N n_N}{m_\pi^2 f_\pi^2}\right) \left| \cos \left ( \frac{a}{2 f_a} \right ) \right| +  \mathcal{O}\left(\left ( \frac{\sigma_N n_N}{m_\pi^2 f_\pi^2} \right )^2 \right)\right\},
\eea
with
\bea
\sigma_N &\equiv& \sum_{q = u,d} m_q \frac{\partial m_N}{\partial m_q},
\eea
where $n_N$ is the number density of nucleons. In this equation, and for the rest of the discussion, we make the simplifying approximation $m_u \approx m_d$.  In vacuum, the axion has a positive mass and is stabilized at the origin.  The situation is very different at finite densities where the coefficient of the cosine can switch sign and the axion is, instead, stabilized around $a/f_a \approx \pi$.  
Because $\epsilon \lesssim 1$, the axion potential can change sign while perturbation theory is still valid.  There is no need to appeal to deconfinement or any other non-perturbative phenomenon that may occur at high densities.  In practice, the neutron star is dense enough that higher order terms might be important. Since $\sigma_N \sim 59$ MeV~\cite{Alarcon:2011zs}, if one takes a solar mass neutron star with a radius of 10 km, then the expansion parameter is 0.4, potentially beyond the regime where the leading order term is an accurate account of the whole correction.\footnote{We do not expect the corrections to change sign as the density of neutrons continue to increase, as in the large density limit, QCD deconfines and the QCD contribution to the axion potential vanish (see review~\cite{Alford:2001dt} and references within).} Therefore, throughout this paper, we will only consider parameter space where $\epsilon \leq 0.1$.

The change in sign of the axion potential at high densities allows the axion to be sourced by objects with near nuclear densities, and if we allow for an even more tuned axion mass, less dense objects as well.  There are two objects with such densities in the universe: neutron stars and nuclei.  An energetics argument shows why neutron stars can source axions while nuclei do not.  To see if the $a = 0$ solution is unstable to perturbations, we can simply compare the gradient energy required to move the axion away from zero ($f_a^2/r^2$) with the gain in potential energy $m_\pi^2 f_\pi^2 (\epsilon - \frac{\sigma_N n_N}{m_\pi^2 f_\pi^2})$.
Only when the gain in potential energy outweighs the gradient energy does the axion get sourced.  We estimate that the axion is sourced if the neutron star is larger than
\bea \label{Eq: rcrit}
r_\text{crit} \gtrsim \frac{1}{m_T}, \qquad m_T = m_\pi f_\pi \frac{\sqrt{\frac{\sigma_N n_N}{m_\pi^2 f_\pi^2} - \epsilon}}{2 f_a},
\eea
where $m_T$  is the tachyonic mass of the axion inside the neutron star or nuclei.  Because we have restricted ourselves to the case where $\epsilon \leq 0.1$, we are in the situation where $m_T \gtrsim m_a$ in a neutron star.   Due to supernova bounds, we are required to have $r_\text{crit} \gtrsim 10^{-8}$ km $= 10^{10}$ fm.  Thus nuclei are too small to source the axion, while neutron stars can do so with ease.  This allows one to avoid fifth force constraints as they are all experiments performed using objects which cannot source the axion in this way.  

The more weakly coupled the axion is (the larger $f_a$), the more it is displaced in field space.  The greater the displacement, the larger the gradient energy is.
Thus we are surprised to find that the more weakly coupled the axion, the stronger the force it mediates!  Conversely, using equation~\ref{Eq: rcrit} we see that the more weakly coupled the axion, the larger and denser the object needs to be to source the axion.
In section~\ref{Sec: massless force}, we will show that under certain assumptions, analytic expressions for the force can be derived.  The force is a standard Yukawa force with an effective charge of order $4 \pi f_a r_\text{NS}$ where $r_\text{NS}$ is the radius of the neutron star. We will also show that neutron stars can have one of two opposite axion field values on the surface. These neutron stars are analogous to conductors with constant positive and negative potentials in classical electromagnetism, and can have both attractive and repulsive axion forces between them, depending on whether the two neutron stars in a binary have the same or opposite field values, respectively. This is in contrast to ordinary scalars and pseudo-scalars which usually mediate attractive forces.

There is an interesting numerical coincidence when considering axion forces.  In order for the force to be observable, we will be considering axion forces comparable to gravity.  This means that $f_a \sim M_p$.  Amusingly, for this value of the mass, we find that $1/m_a \sim 10$ km$/\sqrt{\epsilon}$.  Thus the force is naturally already of order 10 km or longer.  The less tuned the axion mass, the closer the inverse of the mass is to being around 10 km.  This coincidence is also a double-edged sword as the requirement that neutron stars be larger than the critical radius implies that the axion force also cannot be much stronger than gravity, as 10 km is already the size of the neutron star.  

Having a new force between neutrons stars is exciting because Advanced LIGO has the potential to detect neutron star mergers and can probe new forces between neutron stars.  
The fact that the new force has an associated length scale means that at distances $\sim 1/m_a$, the frequency of orbit and therefore the frequency of the emitted gravitational waves will change.\footnote{This requires the two neutron stars in the binary to not be identical.}
When the frequency of rotation becomes larger than $m_a$, then scalar Larmor radiation can occur and therefore the rate of change of the frequency of the gravitational waves will change. These two physical effects will be the main observable consequence at Advanced LIGO. A more drastic effect occurs when the axion force is stronger than gravity and is repulsive.  In this case, the neutron stars do not merge and instead come to rest at a fixed distance apart from each other. The waveform of the gravitational wave emitted in this case has features very different from a merger waveform.

This paper is organized as follows. Section~\ref{Sec: massless force} demonstrates that dense objects can source axions, and discusses the analytic calculation of the force between two neutron stars in various limits.
Section~\ref{Sec: LIGO} provides barebones estimates of how a new force between neutron stars would manifest itself at Advanced LIGO.
Section~\ref{Sec:nuclear} supplies some constraints on these models coming from other processes.
Finally, we conclude in section~\ref{Sec: conclusion}.

\section{Force between neutron stars} \label{Sec: massless force}

\begin{figure}
  \centering
  \includegraphics[width=0.5\textwidth]{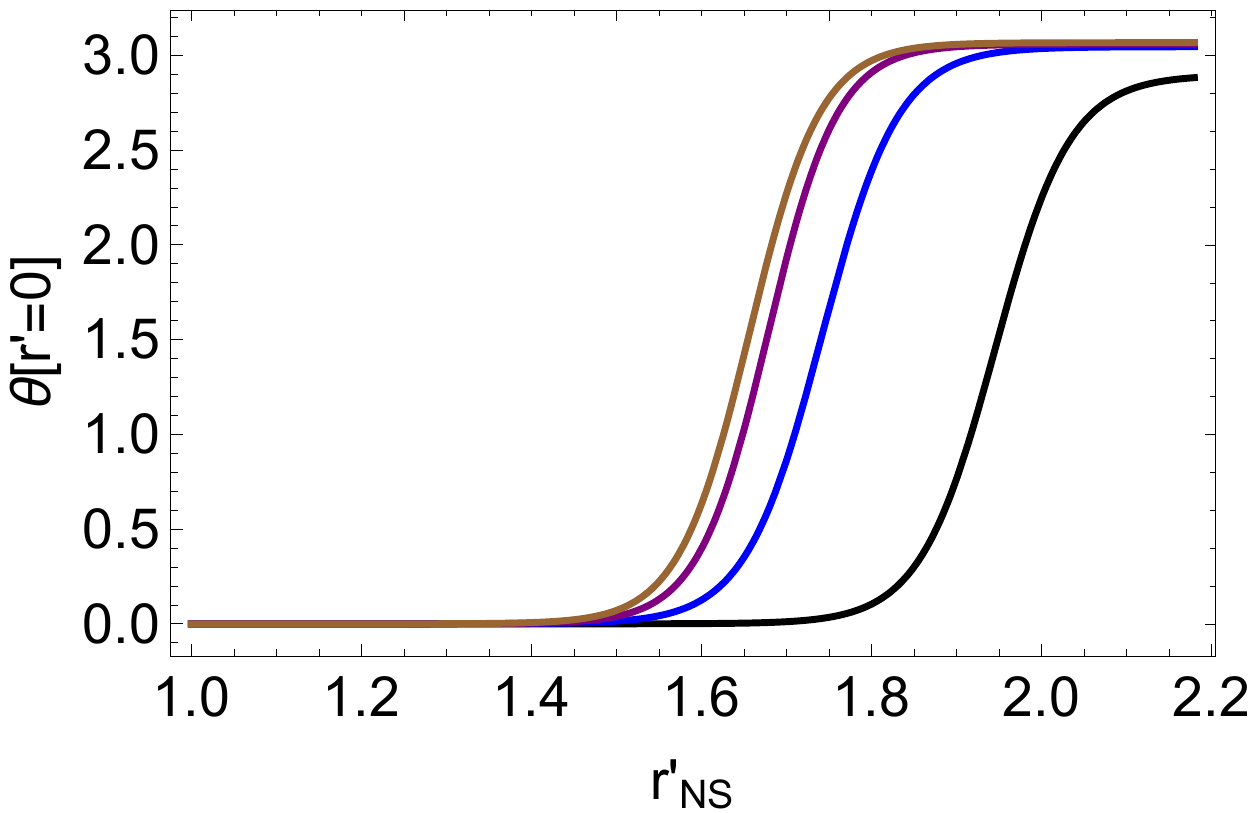}
  \caption{
       Value of $\theta = a/f_a$ at the center of the neutron star as a function of $r'_\text{NS}$.  The four curves correspond to $m_a^2/m_T^2 = $ 0.1 (black), 0.01 (blue), 0.001 (purple) and 0 (brown) from right to left.  An initial profile was assumed and then time evolved with friction towards the stable solution.  The resulting data points were fitted to a smooth curve.  There is clearly a phase transition where only at a particular radius does the neutron star start to source the axion. 
    \label{Fig: source}}
\end{figure}

In this section, we will calculate the strength of the axion force between two neutron stars compared to gravity.  The first step is to consider the case of a single neutron star in isolation.
For simplicity, we model the neutron star as a constant density object such that the equation of motion for the axion is
\bea
\label{Eq: diff eq}
\Box' \frac{\theta}{2} = \left (-1 + \Theta (r' - r'_\text{NS}) (1 + \frac{m_a^2}{m_T^2} ) \right ) \sin \left ( \frac{\theta}{2} \right ) \, \text{sgn} \left\{\cos \left( \frac{\theta}{2} \right )\right\}, \\
\theta \equiv \frac{a}{f_a}, \qquad  r' \equiv r \, m_T, \qquad \qquad \qquad \qquad \nonumber
\eea
where $m_a = m_\pi f_\pi \sqrt{\epsilon}/2 f_a$ is the mass of the axion outside the neutron star, $m_T = m_\pi f_\pi \sqrt{\frac{\sigma_N n_N}{m_\pi^2 f_\pi^2} - \epsilon}/2 f_a$ is the tachyonic mass of the axion inside the neutron star, $r_\text{NS}$ is the radius of the neutron star and the sign function is needed to implement the fact that the potential contains an absolute value.

As argued in the Introduction, comparing the gradient energy with the potential energy shows that only when $r_\text{NS} \gtrsim 1/m_T$ does the neutron star source an axion field.  As seen in figure~\ref{Fig: source}, a phase transition around $r'_\text{NS} \sim 1$ occurs rather robustly and is only mildly sensitive to the value of $m_a^2/m_T^2$.  Small values of $r'_\text{NS}$ do not source the axion while large values do. The profile shown in figure~\ref{Fig: source} is insensitive to the axion profile assumed as the initial condition of time evolution.

Outside the neutron star, the axion potential is roughly $V \approx m_a^2 a^2/2$.  In this limit, one can see that the only solution for the fall off of the axion field is $a = q_\text{eff} e^{-m_a r}/r$.  The axion falls off like any other massive Yukawa interaction with an effective coupling $q_\text{eff} \sim 4 \pi f_a r_\text{NS}$.  Thus one can view a neutron star as a boundary condition where the vacuum expectation value (vev) of the axion is set to be $\sim f_a$.  

There are several cases where explicit analytic expressions for the forces between neutron stars can be calculated.  The simplest example of such an effect is in the limit where $1/m_a \gg D \gg r_{\rm NS}\gg 1/m_T$, where $D$ is the distance between the neutron stars.  The axion mass can be neglected and the neutron stars can be treated as point-like axion source. In this limit, the axion field sourced by a pair of neutron stars is
\bea
a = \frac{q_1}{4 \pi | r - r_1 |} + \frac{q_2}{4 \pi | r - r_2 |},
\eea
where the two neutron stars have effective charges $q_{1,2} \sim 4 \pi f_a r_{\rm NS}$, positions $r_{1,2}$ and we have imposed that the axion field falls off to zero at infinity.  The potential energy between the two neutron stars is therefore the potential of two point charges with equivalent charge $q_1$ and $q_2$
\bea
V = - \frac{q_1 q_2}{4 \pi D},
\eea
where $D = | r_1-r_2 |$.  Thus an inverse-square-law force pulls the two neutron stars closer (farther) in the case where $q_1 q_2 >0$ ($q_1 q_2  < 0$).  This result is not surprising as in this limit, the neutron stars behave like standard point sources. 

The force between neutron stars can be attractive or repulsive.  To see this, note that for every solution $\theta_1(r)$, there is a second solution $\theta_2(r) = - \theta_1 (r)$ that also satisfies equations~\ref{Eq: diff eq}.  At long distances, this means that $q_1 = - q_2$, resulting in a repulsive force.

The result can be easily extended to the case where $D \gtrsim 1/m_a \gg r_{\rm NS}\gg 1/m_T$. In this limit, the axion potential outside a neutron star will be approximated as $V_a\approx m_a^2 a^2/2$, which makes the axion field exponentially small at distances $D > 1/m_a$. The potential energy between two neutron stars becomes $V = - q_1 q_2 e^{-m_a D}/4 \pi D$, the standard Yukawa potential. 

\begin{figure}[t]
  \centering
  \includegraphics[width=0.6\textwidth]{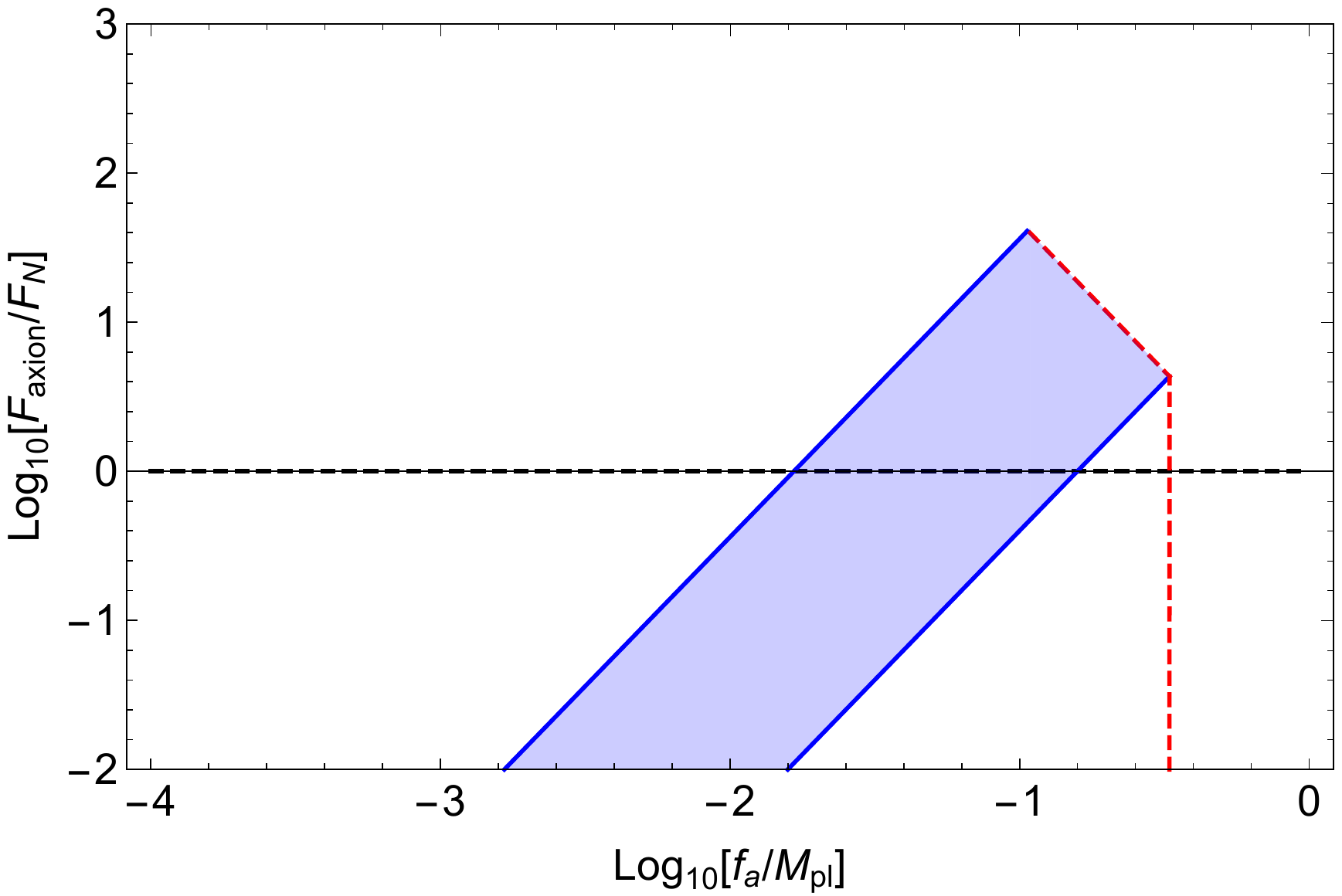}
  \caption{
           The strength of the axion force between two neutron stars compared to the gravitational attraction as a function of the axion decay constant $f_a$ with a vanishing axion mass. The blue shaded region are regions of parameter space that are allowed after taking into account constraints on the neutron star equation of state. The upper and lower blue solid lines come from the maximal and minimal allowed neutron star radius-mass-ratio, respectively. The red dashed line marks the maximal $f_a$ above which the neutron star is not dense enough for our effect to take place.
    \label{Fig:forcefa}}
\end{figure}

The strength of the force between two neutron stars due to axions compared to newtonian gravity is therefore, neglecting the exponential factor,
\bea
\frac{F_{\rm axion}}{F_{\rm N}} \approx \frac{(q r_{\rm NS} f_a)^2}{4 \pi G M_{\rm NS}^2} = \frac{1}{2}\left(\frac{q f_a}{4 \pi M_{\rm pl}}\right)^2 \left(\frac{r_{\rm NS}}{G M_{\rm NS}}\right)^2,
\eea
where $M_{\rm pl}$ is the reduced Planck scale.  The asympotic value of $q/4\pi \sim \pi$ can be found numerically assuming that the neutron star is much larger than the size of the tachyonic mass.  This suggests that, for an ordinary neutron star equation of state, $f_a \approx M_{\rm pl}/10$ would lead to an axion force as strong as gravity, as shown in figure~\ref{Fig:forcefa}. Long range forces between neutron stars can be probed by measurement of neutron star-pulsar and double pulsar binaries. Forces comparable to or stronger than gravity with range as long as roughly a light second are excluded by existing measurement of the Hulse-Taylor binary~\cite{Taylor:1982zz} and PSR J0737-3039~\cite{Burgay:2003jj} (see, for example,~\cite{Will:2001mx} for a detailed review). Forces with a strength as small as $ 10^{-6}$ that of gravity can be probed with current and future measurement of these binary systems, potentially closing the gap between the solar constraint and existing binary system measurement at low masses (see figure~\ref{Fig: nuclear}). Determining the exact constraint one gets from the measurement of these binary systems depends on neutron star equation of states, and requires a dedicated analysis of the orbital evolution, which we leave to future analysis.

\begin{figure}[t]
  \centering
  \includegraphics[width=0.45\textwidth]{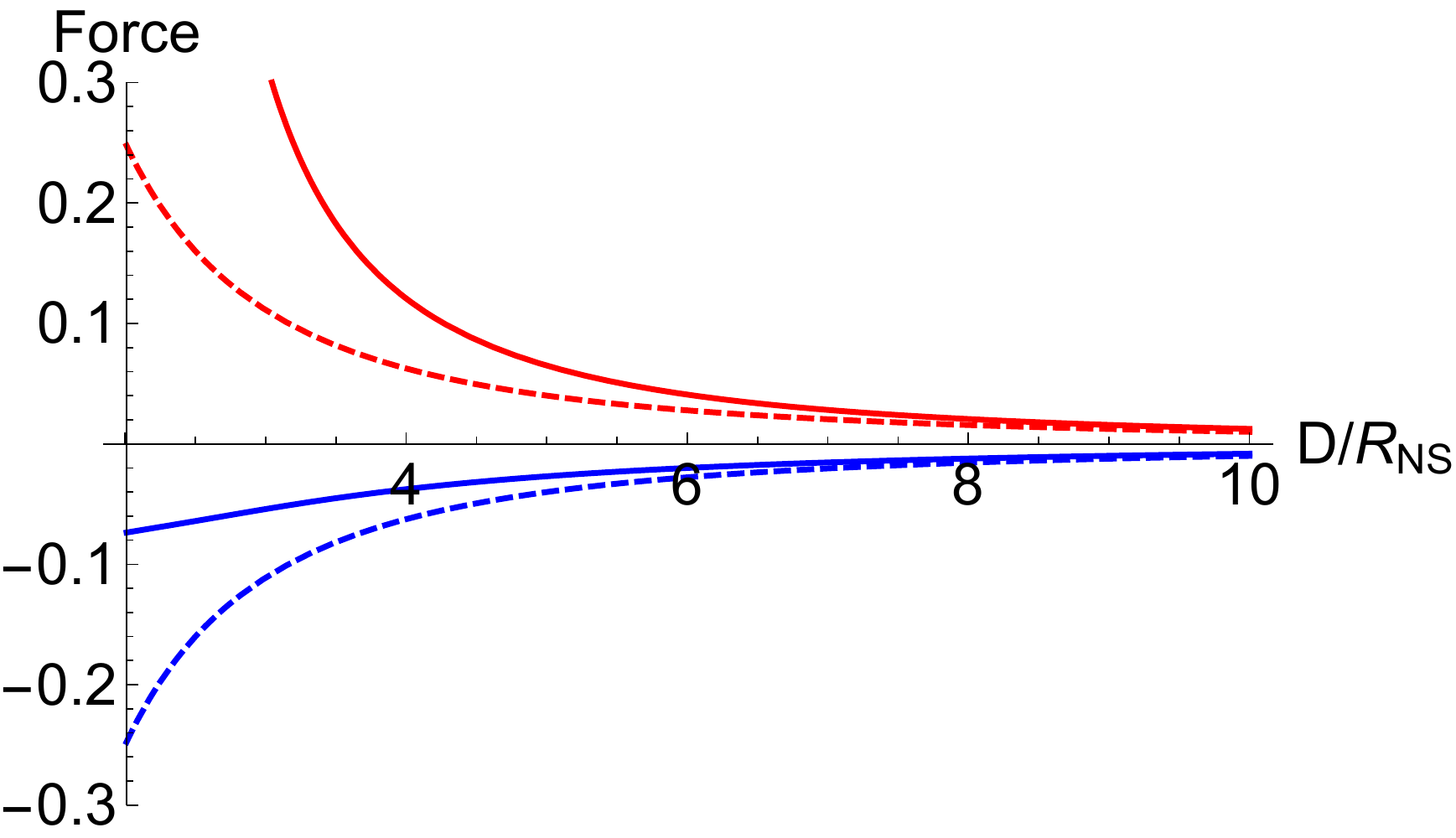}
  \includegraphics[width=0.45\textwidth]{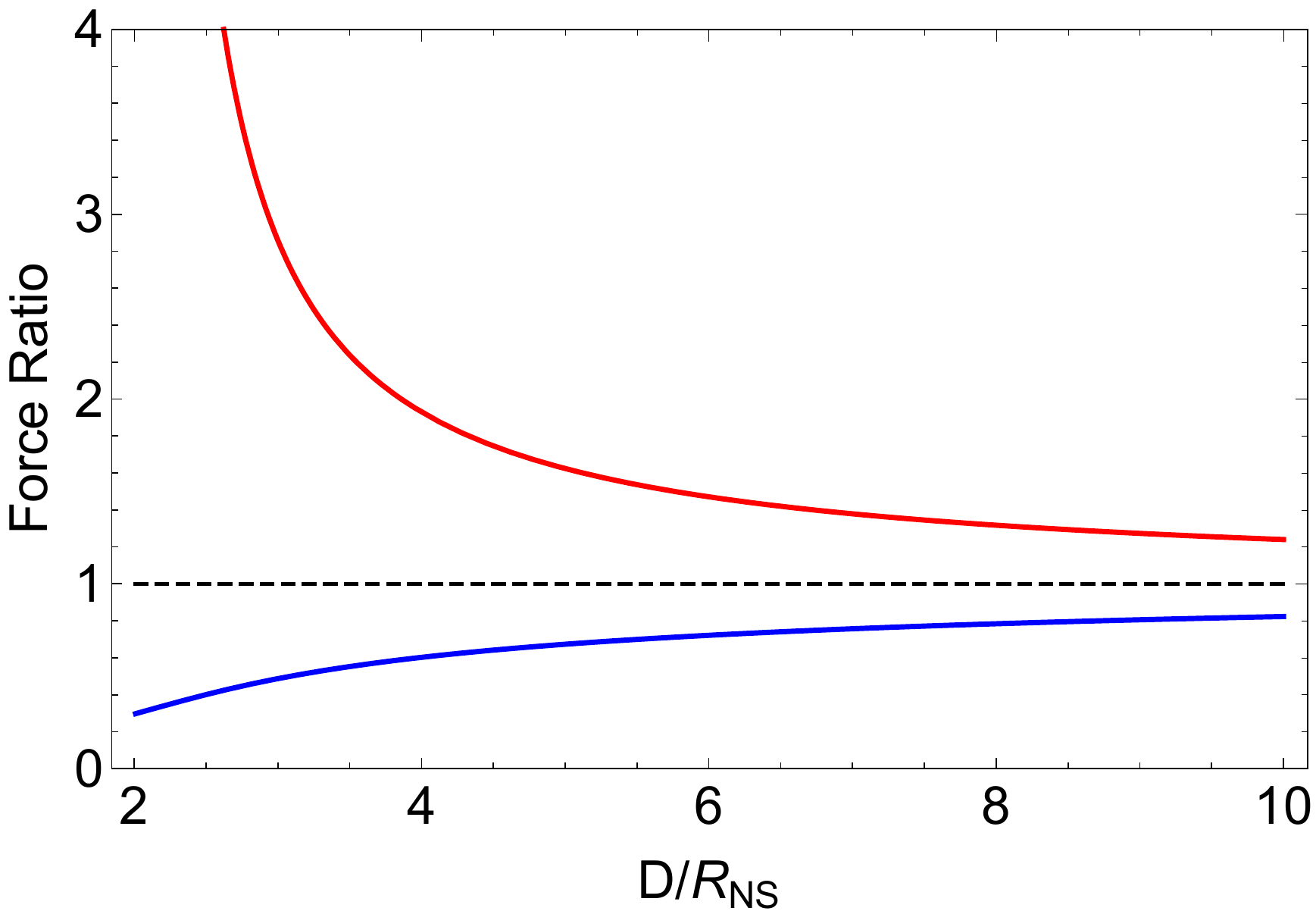}
  \caption{
       (Left Figure) The force (arbitrary units) between the two neutron stars as a function of the distance between them. The blue solid curve shows the attractive force between two neutron stars with the same axion field value inside them ($\theta'(r) = \theta(r)$) while the red solid curve shows the repulsive force between two neutron stars with opposite axion field values ($\theta'(r) = -\theta(r)$). The dashed lines show the same force but with neutron stars treated as point sources. (Right Figure) The ratios between the attractive (blue curve) and repulsive (red curve) forces mediated by the axion to the corresponding forces between point sources.  At short distances, the attractive force is weaker than the forces between point sources while the repulsive force is stronger than the forces between point sources.
    \label{Fig:ForceRatio}}
\end{figure}

The generalization to the case where $1/m_a \gg D \gtrsim r_{\rm NS}\gg 1/m_T$ is less obvious. In this case, the force between two neutron stars is analogous to a calculation of the force between conductors in classical electromagnetism, where image charges are introduced to maintain a spherical equal-potential surface. The method of image charges can also be used to find the potential energy of the system as a function of the distance $d$ between the two neutron stars. When the axion field value in the two neutron stars is the same, as the neutron stars get closer, the image charge required to maintain a constant potential $\theta$ on the surface of a neutron star becomes more negative, and the energy of the system decreases.  In this case, the force between the neutron stars mediated by the axion is attractive. On the other hand, if the axion field values inside the two neutron stars are different, the force between the neutron stars is repulsive. The exact distance dependence of the force is shown in figure~\ref{Fig:ForceRatio}. The repulsive force becomes stronger at shorter distances compared to the force between point-like objects, while the attractive force becomes weaker.

In a neutron star merger event where the force mediated by the axion is weaker than gravity, the additional axion force shows up as an anomalous weakening of the attraction between neutron stars as they approach each other throughout the inspiral phase. The effect of the axion force slowly turns on as the neutron stars inspiral, contrary to the deviations from post-Newtonian results due to distortion of the neutron stars studied in~\cite{Anderson:2007kz}, which will dominate when the neutron stars are less than $\sim 40 \,{\rm km}$ apart (see~\cite{Baiotti:2016qnr} for a detailed review).
The most general case where $1/m_D \sim D \gtrsim r_{\rm NS}\gtrsim 1/m_T$ cannot be treated analytically and will be examined in future work.

\section{Observational consequences for Advanced LIGO} \label{Sec: LIGO}

The axion force between neutron stars discussed in the previous sections can be tested at Advanced LIGO.  In this section we calculate how the signal-to-noise ratio (SNR) of neutron star inspirals depend on the existence of a new force, assuming that precise waveforms of the gravitational wave emission with the axion force will become available.  Numerical work will be important in determining the details, but the qualitative effects can be seen using simple analytic expressions.

Before we get into a detailed calculation of several specific examples, let us first summarize the main changes to the neutron star inspiral that leads to changes to the waveform. The additional attractive or repulsive force mediated by the axion, as well as the radiation of axions, will change the rotational frequency of an inspiral as well as how the rotational frequency changes with time. These changes will lead to changes to the amplitude of the gravitational wave as well as the quality factor of the gravitational wave. Additional scalar Larmor radiation increases the change in frequency and thus decreases the quality factor, making the signal less visible.  The additional force affects both the amplitude and quality factor, and therefore, depending on the strength, range and sign of the force, the final effect might be rather different. In the following, we will use a few special cases to explain how the SNR depends on the strength and range of the force for both attractive and repulsive forces

We can estimate the rough behavior of inspiral phase of the neutron stars by using Newtonian mechanics with gravitational quadropole radiation and scalar Larmor radiation (See appendix~\ref{App: larmor} for a derivation of scalar Larmor radiation), as well as radiation reaction.  We take the inspiral to proceed as circular motion with radiation resulting in a time-dependent radius.  In particular, we solve the system of equations
\bea
\label{eq:radaitionnew}
\frac{dE}{dt } &=& -\frac{32}{5} G \mu^2 D^4 \omega^6 -  \frac{1}{4} \frac{\omega^4 p^2}{6 \pi} (1-\frac{m_a^2}{\omega^2})^{3/2} \Theta(\omega^2 - m_a^2)  \nonumber\\
\frac{d V}{dD} &=& \mu D \omega^2,
\eea
with
\bea
V &=& - \frac{G M_1 M_2}{D} - \frac{q_1 q_2 e^{-m_a D}}{4 \pi D} \nonumber\\ 
E &=& \frac{1}{2} \mu D^2 \omega^2 + V,
\eea
where $\mu = \frac{M_1 M_2}{M_1+M_2}$ is the reduced mass, and $p = q_1 r_1 -q_2 r_2$ is the equivalent of a dipole moment. The orbital frequency $\omega$ and inter neutron star distance $D$ are both time dependent functions. Note that in this calculation, we are using the Yukawa approximation for the force.  This assumption is valid as most of the statistical significance of the inspiral is obtained when $r_{\rm NS}/D \lesssim 0.1$.

From this Newtonian approximation, we can derive the form of the gravitational waves seen by Advanced LIGO.  Ignoring red-shift factors, the gravitational waves have the form
\bea
h_{+}(t) = \frac{4 G \mu \omega^2 D^2}{r} \frac{1 + \cos\theta_i}{2} \cos 2 \omega t,  \qquad h_{\times}(t) = \frac{4 G \mu \omega^2 D^2}{r} \cos \theta_i \sin 2 \omega t ,
\eea
where $\theta_i$ is the inclination of the system relative to the observer (for simplicity we will set it to 0 for the rest of the paper) and r is the distance to the source.  The Fourier transform of the gravitational wave can be calculated using the stationary phase approximation and is
\bea
|\tilde h^2(f)| = |\tilde h_{+}^2(f)| + |\tilde h_{\times}^2(f)| = \frac{4 \pi G^2 \mu^2 r^4 \omega^4}{D^2 \dot \omega} |_{\omega = \pi f}.
\eea
The frequency of the gravitational wave is related to the frequency of orbit by $w = \pi f$, as expected from quadropole radiation.  

Assuming an optimal filter, the total SNR can be calculated to be
\bea
\text{SNR}^2 = 4 \int_0^\infty df \frac{|\tilde h^2(f)|}{S_n(f)},
\eea
where $S_n(f)$ is the one-sided power spectral density.  When estimating the SNR, we use the projected sensitivity of Advanced LIGO~\cite{Aasi:2013wya}.  
In our calculations, we only use the inspiral phase of the neutron star merger as the merger itself requires careful numerical simulation.

\begin{figure}[t]
  \centering
  \includegraphics[width=0.45\textwidth]{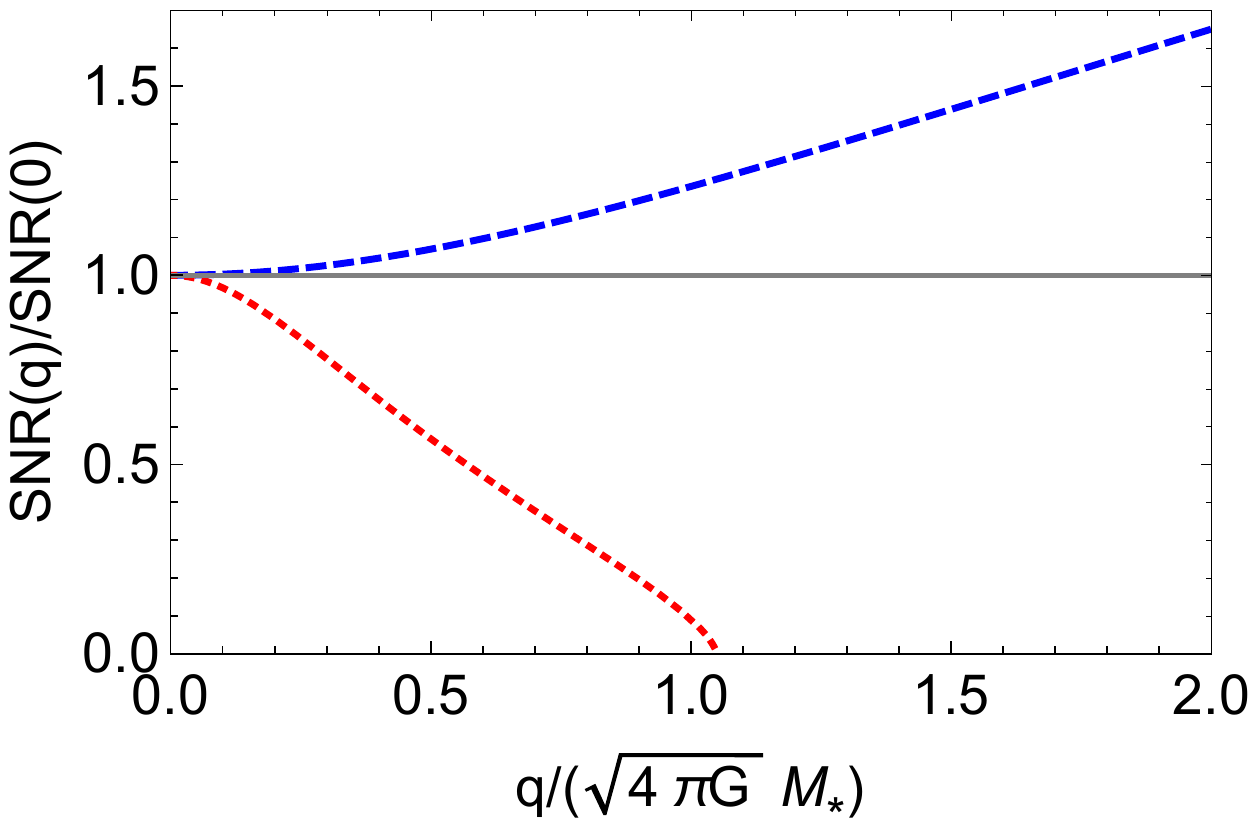}
  \includegraphics[width=0.45\textwidth]{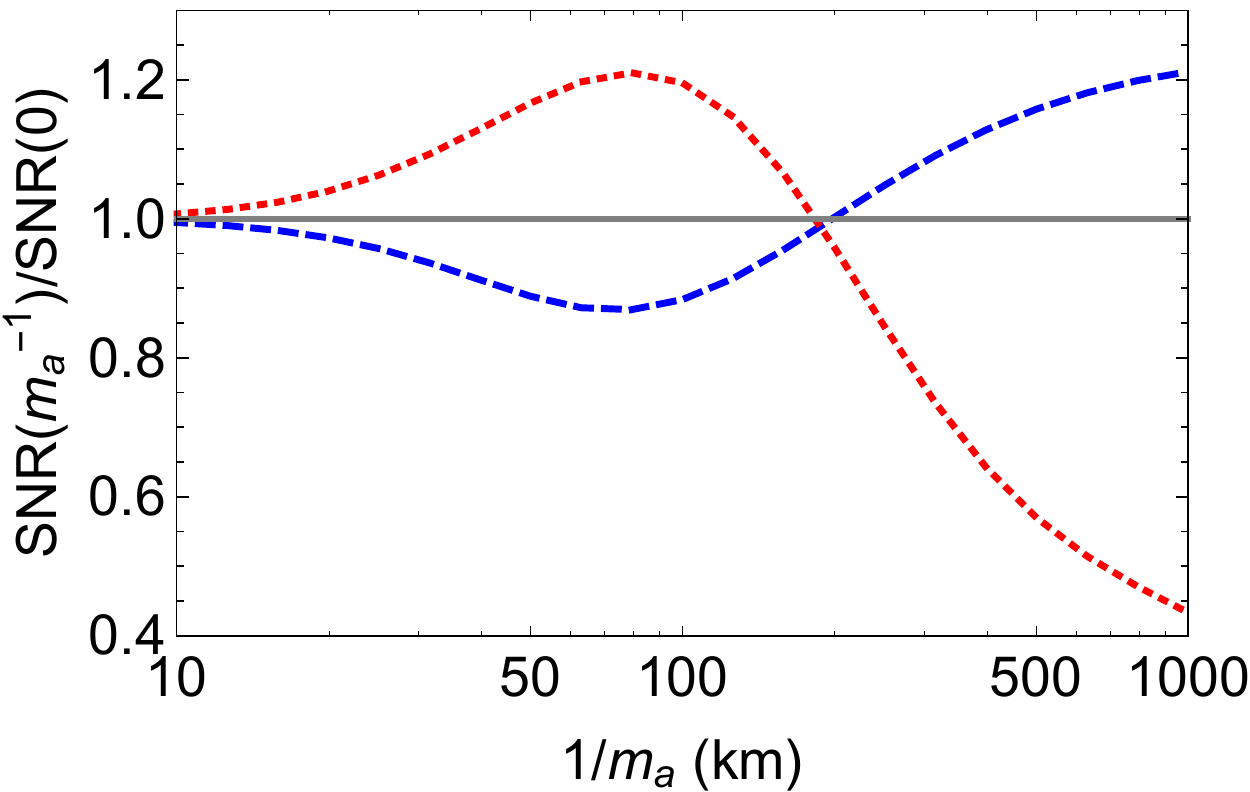}
  \caption{
           (Left Figure) The dependence of the total SNR on the charge q with a massless axion.  The charge of the neutron stars is assumed to be the same while the masses are assumed to be 1 and 1.1 solar masses.  The different masses are chosen to ensure that the scalar dipole moment does not vanish.  The blue dotted curve is a new attractive force while the red dashed curve is a new repulsive force.  We see that repulsive interactions make the neutron stars harder to detect while attractive interactions make them easier to detect.
           (Right Figure) The dependence of the total SNR on mass of the axion.  As before, we choose a 1 and 1.1 solar mass black hole with charges $\pm1$ so that the asymptotic result when $m_a \rightarrow 0$ matches the figure on the left.  As the range of the force approaches to the 100 km scale, the attractive (repulsive) interaction goes from being more (less) visible than normal neutron stars to being less (more) visible.
    \label{Fig: snr}}
\end{figure}

What is of interest is whether the new axion force makes the merger of neutron stars more or less visible.  
Figure~\ref{Fig: snr} shows the total SNR as a function of the effective charge $q$
and axion mass and compares it to the total SNR of the same neutron stars with no charge.

Two competing effects lead to the structure seen in figure~\ref{Fig: snr}.
At a fixed frequency, the SNR of the gravitational wave is proportional to the radius to the fourth and inversely proportional to $\dot \omega$.  If a new attractive (repulsive) force is present, then the radius increases (decreases).  This increase (decrease) in the amplitude of the gravitational waves can be naturally understood as the neutron stars moving faster (slower) than they would have without the new force.  
The new attractive (repulsive) force also decreases (increases) the quality factor $Q \sim w/\sqrt{\dot \omega}$ by increasing (decreasing) the amount of radiation, both gravitational and scalar, being emitted by the system, reducing the amplitude of the gravitational wave.  
When $m_a = 0$, the effect of the change in radius is a larger effect so that an attractive (repulsive) force results in an enhanced (diminished) SNR.  As $m_a$ varies away from zero, scalar Larmor radiation turns off as the mass becomes larger than the orbital frequency (a lesser effect), and the axion force deviates more from a $1/r^2$ form, both making the radius more dependent
on $m_a$ than $\dot \omega$ is.  As a result, when $m_a$ approaches some critical value depending on the charge of the neutron star (see figure~\ref{Fig: snr} for an example), the effects flip sign and an attractive (repulsive) force results in a diminished (enhanced) SNR.

In the case where the axion force is weaker and comparable to gravity, there are two main observable features that distinguish our axion force from general relativity as well as some of its extensions~\cite{Okounkova:2017yby,Endlich:2017tqa,Berti:2015itd} (aside from the fact that such an effect only happens in a neutron star inspiral). Firstly, as the inspiral radius reaches the Compton wavelength of the axion $1/m_a$, the axion force ceases to be exponentially small and significantly change the frequency of the orbital rotation. In particular, in the case of a repulsive force, this might lead to a decrease of orbital frequency over time, which does not happen in general relativity. Secondly, when the orbital frequency becomes comparable to $m_a$, scalar Larmor radiation turns on, which leads to an increase of $\dot \omega$ that cannot be accounted for by the back-reaction of the gravitional wave emitted. 
The merger phase of these events may be even more drastic than the inspiral phase as indicated by the super-emitting binaries of Ref.~\cite{Hanna:2016uhs}.

\begin{figure}[t]
  \centering
  \includegraphics[width=0.45\textwidth]{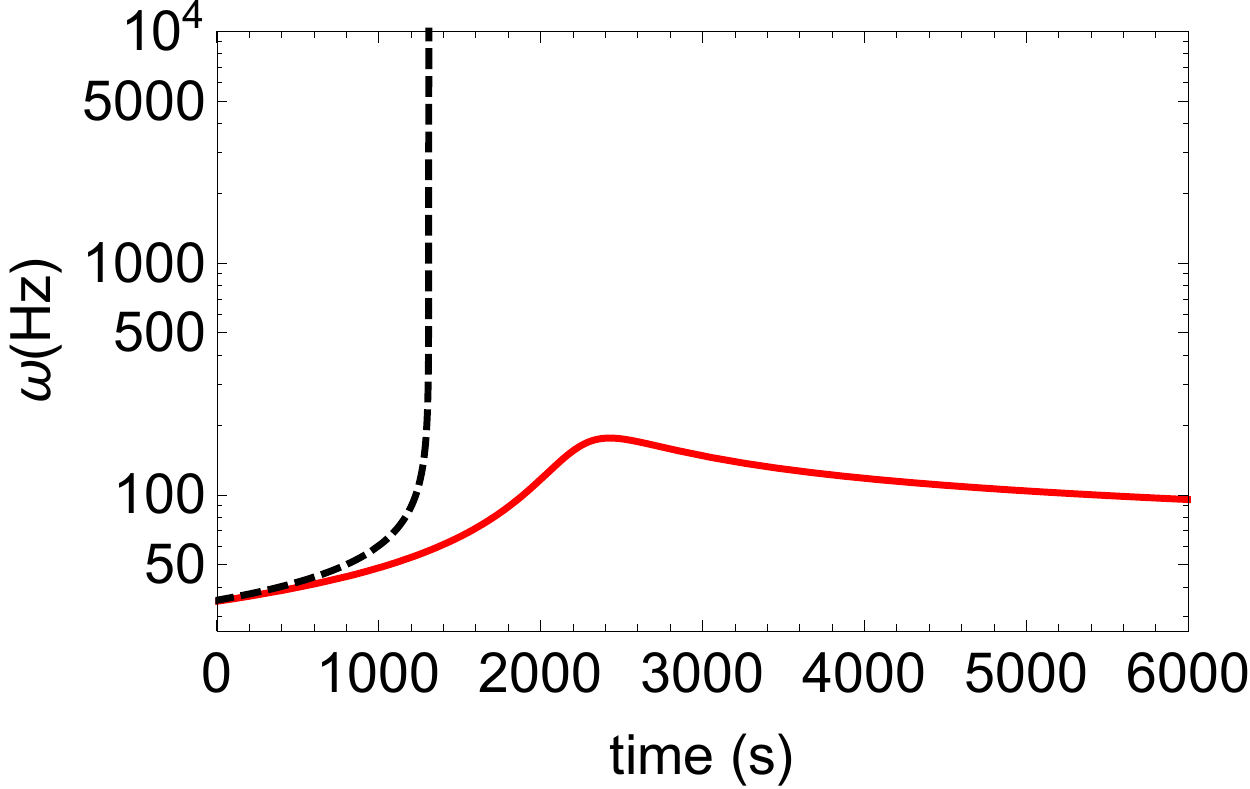}
  \caption{Angular frequency as a function of time for a pair of inspiraling neutron stars where the axion force is stronger than gravity and repulsive.  The black dashed curve shows the result for uncharged neutron stars while red shows the result for neutron stars with a charge.  The frequency increases until the radius of the orbit is $r \sim 1/m_a$, where the repulsive force turns on and the frequency starts to decrease. \label{Fig: repulsive}}
\end{figure}

A more exotic case is when the new repulsive force is stronger than gravity (see figure~\ref{Fig: repulsive} for the evolution of frequency as a function of time).  Rather than merge, the two neutron stars inspiral until $r \sim 1/m_a$ and then reach their equilibrium positions where gravity balances the stronger but Yukawa-suppressed axion force.  In this case, the total SNR can become significantly enhanced as the quality factor $Q$ becomes very large.  At some point in the orbit, $\dot \omega$ goes through zero and $Q$ diverges.  Then 
one must go beyond the gaussian approximation when using the stationary phase approximation (see appendix~\ref{App: quality}). Solving equation~\ref{eq:radaitionnew} numerically gives the naive estimate that the total SNR in this case can be enhanced by as large as $\mathcal{O}(100)$. However, since the GW emitted in this case has very different waveform compared to GW emitted from a neutron star merger, a careful numerical study of the waveform beyond our crude estimate is especially vital in this case in order to compare with data. 

The most exotic case is when the axion mediates a repulsive force that is comparable to and slightly weaker than gravity. In this case, when the two neutron stars become very close, the repulsive force can become stronger than gravity due to deviations from $1/r^2$ forces at short distances (see figure~\ref{Fig:ForceRatio}). Though such a behavior is very unique to our scenario, it is already in the regime where a dedicated numerical analysis is required as the neutron stars are very close to each other. We will leave this case to future analysis.

\section{Constraints from other measurements}\label{Sec:nuclear}

Just as the axion can be sourced by neutron stars, it can also be sourced by other compact stellar objects such as the earth, the sun, red giants and white dwarfs. 
Due to the much smaller density of these objects compared to the neutron star, the axion potential needs to be proportionally more tuned.  However, since we have a good understanding of the composition and spectroscopy of these objects, we can put constraints on a wide range of allowed axion parameter spaces.

Similar to the neutron star case, when the axion is light, the large densities inside a stellar object can also change the sign of the axion potential and source an axion field with $\theta \sim \pi$ inside the stellar object and a profile
\begin{equation}
\theta(r > r_{\rm S}) \sim \frac{\pi r_{\rm S}}{r} \exp[- m_a r]
\end{equation}
outside, where $r_{\rm S}$ is the radius.  This relation holds as long as the density and radius of the stellar object obeys
\begin{equation}
\rho_{\rm S} \gtrsim m_a^2 f_a^2,\quad {\rm and}\quad 1/r_S \lesssim \sqrt{\rho_{\rm S}}/f_a .
\end{equation}

\begin{figure}[t]
  \centering
  \includegraphics[width=0.99\textwidth]{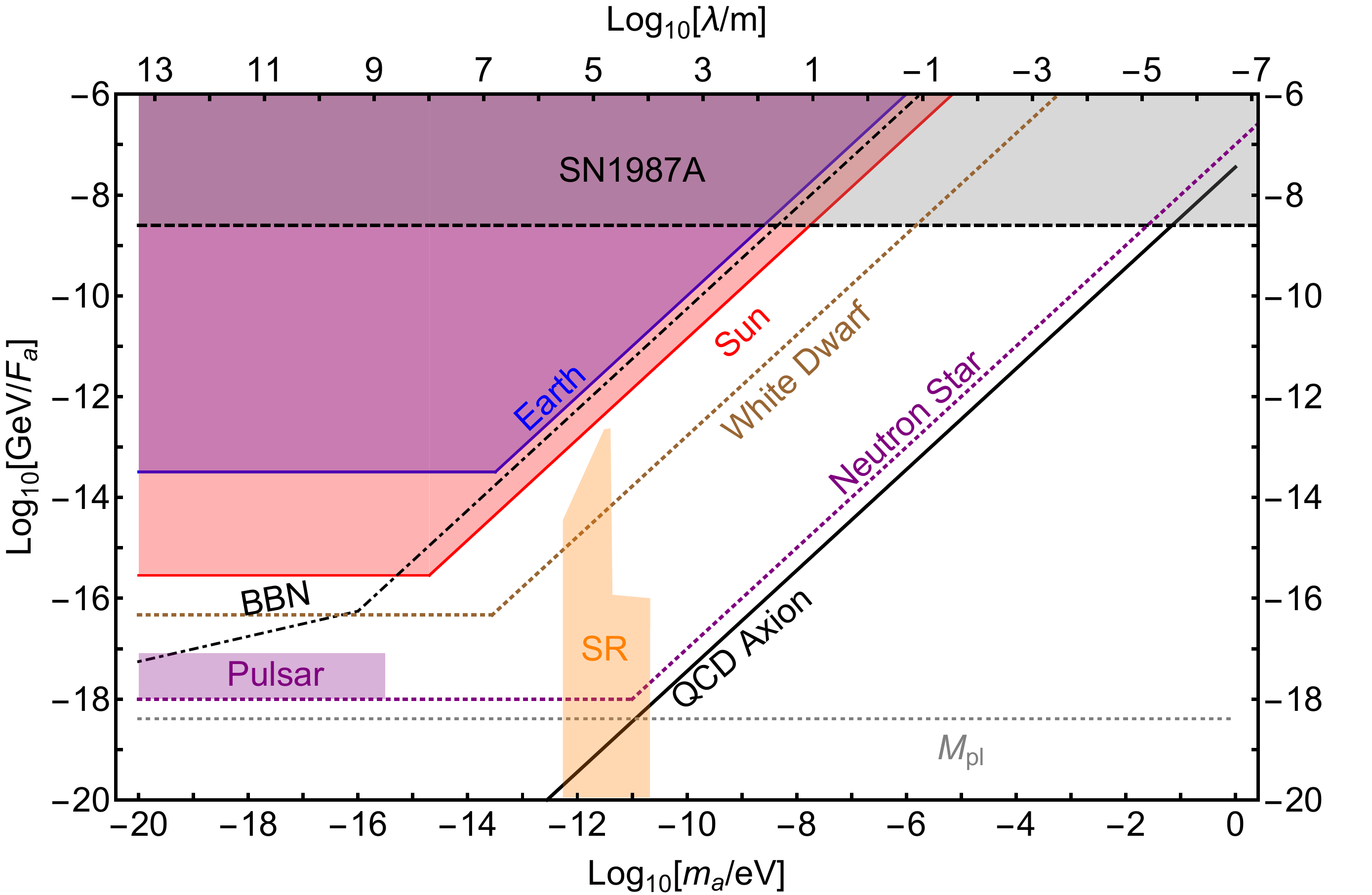}
  \caption{The existing constraints on the axion parameter space due to observations of various stellar objects in the universe. The blue shaded region and the red shaded region are excluded by direct measurements of the earth and the sun, respectively. The region above the dotted brown line and the dotted purple line are parameter spaces that can be probed by indirect measurement of the white dwarf and neutron star, respectively. The exact location of the neutron star line depends on the mass-radius relation of the neutron stars~\cite{Lattimer:2012nd}. The gray shaded region is excluded by SN1987A while the orange shaded region is excluded by blackhole superradiance measurements~\cite{Arvanitaki:2014wva,Arvanitaki:2010sy}. The purple shaded region is an conservative estimate of the exclusion by measurement of orbital decay of binary pulsar systems (axion force not stronger than gravity). The black dot-dashed line shows the constraint from BBN on the axion parameter space if the axion is the dark matter in the universe~\cite{Blum:2014vsa}. The solid black line marks the coupling-mass relation that matches the QCD axion, while the dotted gray line marks the value of the reduced planck scale $M_{\rm pl}$. The region to the left of the orange superradiance constraint can potentially be tested with future LIGO measurements of neutron star mergers. 
    \label{Fig: nuclear}}
\end{figure}

The physical properties of nuclear matter will change dramatically in a medium with a large $\theta$-angle.  An $O(1)$ $\theta$-angle will lead to changes to the masses of the pions, the mass difference between the proton and the neutron, as well as the mass spectrum of stable nuclei.  A simple rule of thumb is that if $\theta = \pi$ then one can simply treat the up quark mass as negative so that the pion mass decreases and the proton neutron mass difference increases (see e.g.~\cite{Ubaldi:2008nf} for details).  These changes can be observed by current and future measurements of stellar objects with various densities and sizes.  The constraints are summarized in figure~\ref{Fig: nuclear} and discussed below.

\subsection{Direct observable consequences}

If the axion is sourced on earth or by a nearby stellar object, then direct observables on earth can be used to exclude it. QCD theta angle that is not extremely close to $0$ or $\pm \pi$ is excluded by various measurements of eDMs. A QCD theta angle of $\pi$ instead of $0$ on earth is excluded, for example, by measurements of the pion masses.  In the chiral Lagrangian with the limit $m_s \gg m_d$ and $m_u$, it can be shown that various sums and differences of pion masses are proportional to $m_u$ so that its sign can be determined to be positive.  Lattice simulations at $\theta = \pi$ would be needed to show that higher dimensional operators do not spoil this argument.
These simulations do not exist due to technical difficulties; however, it seems reasonable to believe that higher dimensional operators are suppressed by appropriate powers of $4 \pi$ as they are around $\theta =0$. Similarly, measurements of neutron proton mass difference, as well as masses of various nucleus, disfavors $\theta \sim \pi$ on earth. Moreover, we have done many dedicated experiments searching for nuclear eDM. The null result from these experiments constrains the QCD $\theta$-angle on earth to be within $10^{-10}$ of either $0$ or $\pi$~\cite{Baker:2006ts}.  If the earth did source the axion, finite radius corrections would result in deviations from $\theta = \pi$ larger than $10^{-10}$.

Similarly, a theta angle of $\mathcal{O}(1)$ is excluded inside the solar core.  The main factor here is that the proton-neutron mass difference will increase by $\sim 10$ MeV if $\theta \simeq \pi$.  If this were the case, then the proton-proton chain reaction in the sun would not be as efficient due to the inability to form deuterium inside the solar core.  Additionally, the Borexino experiment sees the neutrino line coming from inside the sun with an energy $\sim 862$ keV due to the Be$^7$-Li$^7$ mass difference~\cite{Bellini:2013lnn}.  If the proton-neutron mass difference increased by 10 MeV, this reaction would not occur as Li$^7$ would be heavier than Be$^7$ when $\theta \approx 1$ inside the sun. 

\subsection{Indirect observable consequences}

In the earth and sun, we can directly measure the properties of the nucleus through pion masses, eDMs or the neutrino spectrum from beta decay, respectively. For most stellar objects in the universe, however, there are no direct measurements of their nuclear properties.  However, we can infer from indirect observations of their surroundings that the nuclear properties cannot be very different.  

As an example, consider red giants, which are responsible for most of the carbon in the universe.  Because carbon is created resonantly through the triple alpha process, varying $\theta$ can vary the amount of carbon in the universe.  
Requiring that carbon abundance of the universe was reproduced excludes $\theta \gtrsim 10^{-3}$ in red giants~\cite{Ubaldi:2008nf}.

Another type of indirect measurements appertains the X-ray emissions from the surroundings of compact stellar objects. Accreting white dwarfs and neutron stars have an X-ray emission spectrum that reveals the chemical composition of the matter around the respective stellar object. The existence of Oxygen, Nitrogen, Neon and, most importantly, Iron on the surface of white dwarfs and neutron stars can be established through measurements of the X-ray emission of their corresponding ion (see ~\cite{Mukai:2017qww} and reference within). An Iron-rich environment is in good agreement with the common lore that $\mathrm{Fe}^{56}$ is the most stable nucleus when $\theta$ is small~\cite{Audi:2003zz}. However, when the $\theta \simeq \pi$, mostly due to a much larger neutron-proton mass difference~\cite{Ubaldi:2008nf}, $\mathrm{Co}^{56}$ will be lighter than $\mathrm{Fe}^{56}$ by $\sim 5$ MeV and $\mathrm{Fe}^{56}$, if once abundant, would have decayed to $\mathrm{Co}^{56}$ in roughly a few hours through $\beta$-decay. The observation of $\mathrm{Fe} \, \mathrm{K}\alpha$ line combined with the non-observation of the corresponding $\mathrm{Co}$ X-ray emission would suggest that $\theta$ cannot be $O(1)$ in the surroundings of white dwarfs and neutron stars in an X-ray binary.

It is clear that these consequences from indirect observations of X-ray emission of either the white dwarfs or the neutron stars require a good understanding of nuclear properties when $\theta \sim 1$, atomic-nuclear reactions in extreme environments near a neutron star, as well as the density profiles near white dwarfs or the neutron stars.
Dedicated X-ray measurement, as well as more careful theoretical studies, are required to establish a vanishing $\theta$ angle near either white dwarfs or neutron stars, which is beyond the scope of this paper. 

It should be noted that the axion discussed in this paper can also influence the evolution of stellar objects. The axion field profile evolves during the formation of compact objects, e.g. white dwarfs and neutron stars, and might change their evolutionary history  thereby changing their number and mass distributions.  There may even be potentially observable consequences during, in particular, supernova explosions. The understanding of the evolution of these systems require dedicated analyses and simulations, which we leave to future work.

\subsection{Cosmological considerations}

In this subsection, we will discuss the cosmological evolution of the light axion. The available parameter space for the axion strongly depends on whether the non-QCD part of the axion potential was present during the early universe evolution. If the potential was always present during the evolution of the early universe, the axion will roll down the potential at high temperature, resulting in $\theta = \pi$ when the QCD contribution to the potential turns on as the universe cools down. In this case, cosmological considerations would require a reheating temperature $T_{\rm RH} \lesssim 1 \,{\rm GeV}$ so as to avoid overclosing the universe. However, if the non-QCD part of the potential was not yet on, or only partially on, during the QCD phase transition, then $\theta = 0$ can potentially always be a minimum and there is no constraint on the viable reheating temperature.\footnote{If the sector that gives rise to the non-QCD part of the potential has similar matter content to the SM, it is very likely that their contribution to the potential turned on just as the QCD contributions were also turning turns on.  Thus $\theta=0$ would always have been a minimum of the potential.}

Similarly, measurements of Big Bang Nucleosynthesis (BBN) might put a constraint if the non-QCD part of the potential was present during BBN. The nuclear density during BBN is
\bea
n_{\rm nuclear} \sim \frac{n_{\rm nuclear}}{n_{\gamma}} T_{\rm BBN}^3 = 10^{-10} {\rm MeV}^3.
\eea
Such a nuclear density is about two order of magnitude smaller than the density of the solar core. Therefore, the constraints from BBN would be at best comparable to or weaker than the constraint from the sun if the non-QCD potential was always present. 

\section{Conclusions and future directions} \label{Sec: conclusion}

In this paper, we have shown how the axion can mediate a new long range force between neutron stars that can be as strong as gravity, and therefore can be tested with Advanced LIGO and future gravitational wave detectors. 
 
The axion force considered in this paper evades standard fifth force constraints as it is sourced only by neutron stars, and similar large objects with high density.  A new force between neutron stars is exciting because the discovery of black hole mergers is ushering in a new era of data and neutron star mergers will likely be discovered as well by Advanced LIGO.

The axion force has multiple surprising features.  The first is that the more weakly coupled the axion, the stronger the force it mediates between
neutron stars.  The second is that this force can be either attractive or repulsive, which is not expected since scalars and pseudo-scalars typically mediate attractive interactions.  Finally, when the finite size of the neutron star is taken into account, the repulsive force grows faster than $1/r^2$ so that even a force slightly weaker than gravity at long distances might eventually cause the neutron stars to repel each other at short distances.

As a result of the above mentioned features of the axion force, there are many striking signatures that can be searched in Advanced LIGO and future gravitational detectors. Axion force and radiation can change both the amplitude and the quality factor of the gravitational radiation, leading to significant changes to the GW waveforms. Moreover, in cases where a repulsive axion force is stronger than gravity at short distances, neutron stars would not be able to merge even after long periods of inspiral and lead to gravitational waves with qualitatively different waveforms.

We also considered several other large and dense objects that can source the axion. Direct observations of stellar objects close by, e.g.\ the earth and the sun, excludes large part of the axion parameter space. Much denser objects, for example, white dwarfs and neutron stars, might provide valuable new information about the axion parameter space.

The behavior of the axions discussed in this paper is similar to the case of spontaneous scalarization~\cite{Damour:1993hw}.  In brief, spontaneous scalarization is a similar effect that occurs with a dilaton-like scalar.  Due to the coupling of the dilaton squared to $T^\mu_\mu$, if the dilaton has been tuned to be light, then finite density effects can push the dilaton away from the origin resulting in a force between neutron stars.  The salient difference between the dilaton and axion models is that the axion model is significantly less tuned. Density plays a critical role in other forces such as Chameleons~\cite{Khoury:2003rn} and Symmetrons~\cite{Hinterbichler:2010es}.  In these other examples density serves to screen the force so as to avoid fifth force constraints.

Finally, there are many future directions.  Numerical simulations are clearly needed to search for our effect at Advanced LIGO and future gravitational wave detectors, in particular for regions of parameter space which are not under analytic control. In this paper, we have already shown how various regions of the axion parameter space have already been excluded by measurement of particle properties, as well as optical and neutrino signatures. It would be interesting if this could be extended to denser objects like the white dwarf and neutron stars, leading to new and exciting approaches towards discovering the axion. 

\section*{Acknowledgements}

The authors thank Shamit Kachru for collaboration in the early stages of the project. We thank Nima Arkani-Hamed, Asimina Arvanitaki and Matthew Johnson for many interesting discussions. We also thank Masha Baryakhtar, Robert Lasenby, Gustavo Marques-Tavares and Matthew Johnson for useful discussions and for comments on the draft. A.H. and J.H. are supported by NSF Grant PHYS-1316699. A.H. is also supported by the DOE Grant DE- SC0012012. This research was supported in part by Perimeter Institute for Theoretical Physics. Research at Perimeter Institute is supported by the Government of Canada through the Department of Innovation, Science and Economic Development Canada and by the Province of Ontario through the Ministry of Research, Innovation and Science.

\appendix

\section{Axion potential at finite density} \label{App: finite}

We briefly review the results of~\cite{Cohen:1991nk}.  It is straightforward to apply these results to the axion.  We assume that we are dealing with non-relativistic matter made of neutrons and protons.  The axion potential comes from the quark mass terms.  At finite density, the quark mass terms are changed by
\bea
m_u (\langle \overline u u \rangle_{n_N} - \langle \overline u u \rangle_0 ) &=& -m_u ( \langle \frac{\partial H}{\partial m_u} \rangle_{n_N} - \langle \frac{\partial H}{\partial m_u} \rangle_0 ) = - m_u \langle \frac{\partial E}{\partial m_u} \rangle = \sum_{N = n,p} n_N \sigma^u_N , \nonumber \\
 \text{with}\quad \sigma^u_N &\equiv& m_u \frac{\partial m_N}{\partial m_u},
\eea
where $n_{n,p}$ is the number density of neutrons and protons.  An analogous formula holds for down-type quarks.  These results can be applied in a straightforward manner to the axion potential by defining 
\bea
m_{u,d}^\text{eff} = m_{u,d} \left (1 - \frac{\sum \sigma^{u,d}_N n_N}{m_\pi^2 f_\pi^2} \frac{m_u + m_d}{m_{u,d}} \right )
\eea
and using the effective up and down quark masses when calculating the axion mass.  As this formula is rather messy, we give the result in the limit where $m_u = m_d$ in equation~\ref{Eq: finite density}.

\section{Scalar Larmor radiation} \label{App: larmor}

In this section, we calculate the Larmor radiation for the axion for inspirialing neutron stars.  We will approximate the axion potential by just its mass term and again work in the limit where the neutron stars can be approximated by point-like particles.

We are interested in the flux of energy away from a pair of neutron stars moving in a circle.  The power emitted is
\bea
P = r^2 \int d\Omega \int \frac{dt}{T} T^{r0} = r^2 \int d\Omega \int \frac{dt}{T} \partial^r a \partial^0 a
\eea
where we assume that the neutron stars are undergoing motion with a period of T.  As the neutron stars behave as point-like objects, we can find the axion field by treating them as a source ($j = q_\text{eff} \sqrt{\frac{dx^\mu}{dt} \frac{dx_\mu}{dt}} \delta^3(x-x_0)$) so that the axion profile can be solved in fourier space by 
\bea
(k^2 - \omega^2_n + m_a^2) a(\omega_n,k) &=& j(\omega_n,k), \qquad \omega_n = \frac{2\pi n}{T}, \\
a(\omega_n,k) &=& \int \frac{dt}{T} d^3x e^{i (\omega_n t - k \cdot x)} a(t,x).
\eea
The energy being emitted is
\bea
\frac{dP}{d\Omega} = \sum_n  \frac{\omega_n k_n}{16 \pi^2} |j(\omega_n, k_n \, \hat r)|^2 \Theta(\omega_n^2 - m_a^2), \qquad k_n^2 = \omega_n^2 - m_a^2,
\eea
where $\Theta$ is the Heaviside theta function.  Doing the integral gives the result
\bea
P = \frac{1}{4} \frac{\omega^4 p^2}{6 \pi} (1-\frac{m_a^2}{\omega^2})^{3/2} \Theta(\omega^2 - m_a^2) , \qquad p = q_1 r_1 - q_2 r_2.
\eea
In the limit of a massless axion, this is a factor of four smaller than the standard expression for gauge bosons.

\section{Fourier transform of the gravitatioanl wave} \label{App: quality}

In order to determine the SNR of a gravitational wave, one needs a simple analytic calculation of $|\tilde h^2(f)|$.  As is typical, one takes the Fourier transform of the gravitational wave using the stationary phase approximation.
We are most interested in the case where $\dot \omega$ can go through zero so that higher-order corrections must be included.  Taking the gravitational wave to be $h(t) = A(t) \cos 2 \omega(t) t$, we have the Fourier transform
\bea
\tilde h(f) \approx \int dt \frac{A(t)}{2} e^{i (2 \pi f - 2  \omega(t) ) t} \approx \frac{A(t_\star)}{2} \int dt e^{- i \dot \omega(t_\star) t^2} \approx  \frac{A(t_\star)}{2}  \frac{Q_1}{\omega(t_\star)} , \nonumber
\eea
with
\bea
\frac{Q_1}{\omega(t_\star)} = \sqrt{\frac{\pi}{\dot \omega(t_\star)}} , \nonumber
\eea
where one is evaluating everything around the point in time where $\pi f = \omega(t_\star)$.  In this result, we have utilized the fact that $\log A(t)$ is a slowly changing function of time.  
As shown in section~\ref{Sec: LIGO}, if the repulsive force is stronger than gravity, $\dot \omega$ goes through zero and one must series expand $\omega(t)$ to higher order such that 
\bea
\tilde h(f) \approx \int dt \frac{A(t)}{2} e^{i (2 \pi f - 2  \omega(t) ) t} \approx \frac{A(t_\star)}{2} \int dt e^{- i \frac{\ddot \omega(t_\star) t^3}{2}} \approx  \frac{A(t_\star)}{2}  \frac{Q_2}{\omega(t_\star)} , 
\eea
with
\bea
\frac{Q_2}{\omega(t_\star)}  = \frac{2 \pi}{|\Gamma(-1/3)| (\ddot \omega(t_\star) )^{1/3}} . \nonumber
\eea
In order to interpolate smoothly between these two quality factors as $\dot \omega$ becomes small, we use the approximation that
\bea
\tilde h(f) = \frac{A(t_\star)}{2}  \frac{Q}{\omega(t_\star)} , \qquad \frac{1}{Q} = \frac{1}{Q_1} + \frac{1}{Q_2}.
\eea

\bibliography{reference}
\bibliographystyle{JHEP}

\end{document}